\documentstyle[preprint,aps]{revtex} 

\begin{document}

\draft


\title{ Quantum Holonomy in Three-dimensional 
General Covariant Field Theory and Link Invariant} 

\author{ W.F. Chen\renewcommand{\thefootnote}{*}\footnote{
Also at ICSC-World Laboratory, Switzerland}}
\address{ Helsinki Institute of Physics 
P.O. Box 9 (Siltavuorenpenger 20 C), FIN-00014, Helsinki, Finland}

\author{H.C. Lee}
\address{Department of Physics, National Central University, 
Chungli, Taiwan 320, ROC}

\author{Z.Y. Zhu}
\address{Institute of Theoretical Physics, Academia Sinica, 
P. O. Box 2735, Beijing, 100080, China}

\maketitle
                                                        
\begin{abstract}
We consider quantum holonomy  of 
some three-dimensional general
 covariant non-Abelian field theory in Landau gauge  and confirm a
previous result partially proven. We show that  
quantum holonomy retains metric independence 
after explicit gauge fixing and hence possesses the 
topological property of a link invariant.  We examine the  
generalized quantum holonomy defined on a multi-component 
link and discuss its relation to a polynomial for the link. 
\end{abstract}

\vspace{2ex}

PACS number(s): 11.15.Tk, 03.65.Fd, 02.40.Pc

\vspace{3ex}

Some time ago it was shown that quantum holonomy
in a three-dimensional general covariant non-Abelian gauge
field theory possesses topological information of the link on which the 
holonomy operator is defined$\cite{lz}$.   The quantum holonomy 
operator was shown to be a central element of the gauge group so that, 
in a given representation of the gauge group, it is a matrix that commutes 
with the matrix representations of all other operators in the group.  
In an irreducible representation, it is proportional 
to the identity matrix.    
Quantum holonomy should therefore in general 
have more information on the link invariant than the quantum Wilson
loop which, for the $SU(2)$ Chern-Simons quantum field theory,  was shown by 
Witten$\cite{wit}$  to yield the Jones polynomial$\cite{jones}$. 
Horne$\cite{ho}$  extended Witten's result to some other Lie groups. 
The difference between quantum holonomy and the 
Wilson loop becomes apparent in the $SU(N|N)$ Chern-Simons theory, 
where the quantum Wilson loop vanishes identically for any link owing 
to the property of super-trace, but the quantum holonomy$\cite{lz}$ 
yields the important Alexander-Conway 
polynomial$\cite{ac,le,at,rs}$.

 However, the argument used in Ref. \cite{lz} was based 
only on the formal properties of the functional integral and 
complications that may arise from the necessity for gauge fixing in any 
actual computation were not taken into consideration. 
In addition, in a case when a  metric is needed for gauge fixing, 
the metric independence of quantum holonomy may be violated.  
Furthermore, in the standard Faddeev-Popov technique used for 
gauge fixing, ghost fields and auxiliary fields that are introduced reduce 
the original local gauge symmetry to BRST symmetry, and it is no longer 
certain the formal arguments and manipulations used in Ref. \cite{lz} 
to derive its results are still valid. 
As well, the case of  the quantum holonomy defined on 
multi-component links was not explicitly considered. 

The main aim of the present work is to clarify these problems.  
We explicitly work in the Landau gauge\footnote{
The reason why we prefer this gauge is because it allows one to avoid 
the infrared divergence in low dimensional gauge theories; for a  
discussion on Chern-Simons theory, see \cite{gmr}.} 
and confirm the results obtained in Ref. \cite{lz} for the case of a 
one-component contour.  
We then show that a quantum holonomy operator 
defined on a $n$-component link, which by construction is a 
tensor product of those operators defined on the components, 
is a central element of the universal enveloping algebra of 
the Lie algebra of the gauge group and, when evaluated on a set of  
$n$ irreducible representations of the gauge group, has a uniquely 
defined eigenvalue that is a polynomial invariant of the link.  

The quantum holonomy is defined as
\begin{eqnarray}
Z[C]{\equiv}\frac{1}{V}
{\int}{\cal D}A\mbox{exp}\left(iS[A]\right) \, f[A,C]\,,\qquad
f[A,C]{\equiv}P\mbox{exp}\left(i{\oint}_{C}A\right)\, ,
\label{eq1}
\end{eqnarray}
where 
where $C$ is a contour in the three-dimensional manifold $M$, 
$S[A]$ is the action of some three-dimensional general covariant
non-Abelian gauge field theory, $P$ means path-ordering 
and $\displaystyle V=\int{\cal D}g$ is 
gauge invariant group volume. Now we choose the Lorentz gauge condition 
\begin{eqnarray}
F[A]={\partial}_{\mu}(\sqrt{-G}G^{\mu\nu}A_{\nu}) =0, 
\qquad G = \mbox{det}(G_{\mu\nu}), 
\label{lorentz}
\end{eqnarray}
where $G_{\mu\nu}$ is the metric of space-time manifold. 
According to standard Faddeev-Popov procedure, we insert the identity
\begin{eqnarray}
1{\equiv}{\Delta}_F[A]{\int}{\cal D}g{\Pi}_x{\delta}(F[A^g(x)])
\end{eqnarray} 
into Eq.(\ref{eq1}) and obtain 
\begin{eqnarray}
Z[C]=\frac{1}{V}{\int}{\cal D}A{\Delta}_F[A]{\int}{\cal D}g
{\Pi}_x{\delta}(F[A^g(x)])\,\mbox{exp}\left(iS[A]\right) \, f[A,C].
\label{eq3}
\end{eqnarray}
Denoting $A^{g}$ as $A$ and replacing the original $A$ by $A^{g^{-1}}$,
 we rewrite Eq.(\ref{eq3}) as follows
\begin{eqnarray}
Z[C]&=&
\frac{1}{V} {\int}{\cal D}g{\cal D}A^{g^{-1}}{\Delta}_F[A^{g^{-1}}]
{\Pi}_x{\delta}(F[A])\,\mbox{exp}\left(iS[A^{g^{-1}}]\right)\,f[A^{g^{-1}},C]
\nonumber\\[2mm]
&=&\frac{1}{V}
{\int}{\cal D}g{\cal D}A{\Delta}_F[A]{\Pi}_x{\delta}(F[A])\,
\mbox{exp}\left(iS[A]\right)\,f[A^{g^{-1}},C]\nonumber\\[2mm]
&=& \frac{1}{V}
{\int}{\cal D}g{\cal D}A{\cal D}B{\cal D}{\bar{c}}{\cal D}
c\,\left\{\mbox{exp}\left[iS[A]-
i{\int}d^3x{\sqrt{-G}G^{\mu\nu}}({\partial}_{\mu}B^aA_{\nu}^a-
{\partial}_{\mu}\bar{c}^aD_{\nu}c^a)\right]\right. \nonumber\\[2mm]
&& \left.{\times}f[A^{g^{-1}},C]\right\}\,
\label{eq4} 
\end{eqnarray}
where we have used the gauge invariance of $S[A]$ and 
$B^a(x), {\bar{c}}^a(x), c^a(x)$ are respectively 
auxiliary fields, ghost and antighost fields.

   We perform the following maneuver on Eq.(\ref{eq4}). Suppose $g^{'}$ 
is a global group element, write $A$ as $A^{g'g'^{-1}}$ and 
rename  $A^{g^{'-1}}$ as $A$ and hence the original $A$ is 
replaced by $A^{g^{'}}$.  We thus obtain, 
\begin{eqnarray}
Z[C]&=& \frac{1}{V}{\int}{\cal D}g
{\cal D}A{\cal D}B{\cal D}{\bar{c}}{\cal D}c\,\left\{
\mbox{exp}\left[iS[A]+
i{\int}d^3x\sqrt{-G}(B^a{\partial}^{\mu}A_{\mu}^a-{\bar{c}}^a
{\partial}_{\mu}D^{\mu}c^a)\right]\right.\nonumber\\[2mm]
&&\left.{\times}f[(A^{g^{-1}})^{g^{'}},C]\right\}\nonumber\\[2mm]
&=& \frac{1}{V}
{\int}{\cal D}g{\cal D}A{\cal D}B{\cal D}{\bar{c}}{\cal D}c\,
\left\{\mbox{exp}\left[iS[A]
+i{\int}d^3x\sqrt{-G}(B^a{\partial}^{\mu}A_{\mu}^a
-{\bar{c}}^a{\partial}_{\mu}D^{\mu}c^a)\right]\right.\nonumber\\[2mm]
&& \left.{\times} {{\Omega}_{g^{'}}}^{-1}
f[A^{g^{-1}},C]{\Omega}_{g^{'}}\right\}\nonumber\\[2mm]
&=&{{\Omega}_{g^{'}}}^{-1}Z[C]{\Omega}_{g^{'}}\, ,
\label{eq5}
\end{eqnarray}
where we have used the fact that all the fields are in the
adjoint representation of gauge group.
Since Eq.(\ref{eq5}) is  true for every global gauge 
transformations, according to 
Schur's lemma, we conclude that when $Z[C]$ is valued in an 
irreducible representation $\rho$ it has the form, 
\begin{eqnarray}
\rho(Z[C]) = F[C]{\bf 1}_{\rho}\, ,
\label{eq6}
\end{eqnarray}
where ${\bf 1}_{\rho}$ is the matrix representation of the identity 
element in ${\rho}$  and $F[C]$, the eigenvalue of $Z[C]$ in $\rho$,  
is a scalar function depending on the contour $C$. 
Eq.(\ref{eq6}) is one of the results obtained in Ref. {\cite{lz}}
without explicit gauge fixing.

In the following, we shall show explicitly the metric 
independence of quantum holonomy.  All fields are valued in the
adjoint representation of gauge group.  As a first step, Eq.(\ref{eq4})
 can be rewritten as follows
\begin{eqnarray}
Z[C]=\frac{1}{V}{\int}{\cal D}g{\cal D}A{\cal D}B{\cal D}
{\bar{c}}{\cal D}c\,\mbox{exp}i\left\{iS[A]
+i{\int}d^3x\sqrt{-G}G^{\mu\nu}{\delta}_B
[{\bar{c}}^a{\partial}_{\mu}
A_{\nu}^a]\right\}f[A^{g^{-1}},C],
\end{eqnarray}
where the BRST transformations are:  
\begin{eqnarray}
{\delta}_BA^a=D_{\mu}c^a\,,\qquad {\delta}_Bc^a=
\frac{1}{2}f^{abc}c^bc^c\,,\qquad
{\delta}_B{\bar{c}}^a=B^a\,,\qquad {\delta}_BB^a=0\,.
\label{eq8}
\end{eqnarray}
Assuming that the functional measures of fields have no 
dependence on the metric,  we have that  
\begin{eqnarray}
&-&\frac{2i}{\sqrt{-G}}\frac{{\delta}Z[C]}{{\delta}G^{\mu\nu}}
=-\frac{2i}{\sqrt{-G}}
\frac{{\delta}F[C]}{{\delta}G^{\mu\nu}}{\bf 1}_{\rho}\nonumber\\[2mm]
&=&-\frac{2i}{\sqrt{-G}}\frac{\delta}{{\delta}G^{\mu\nu}}\left\{
\frac{1}{V}{\int}{\cal D}g{\cal D}A{\cal D}B{\cal D}{\bar{c}}
{\cal D}c\,\mbox{exp}i
\left[iS[A]+i{\int}d^3x\sqrt{-G}G^{\alpha\beta}{\delta}_B\left({\bar{c}}^a
{\partial}_{\alpha}A_{\beta}^a\right)
\right]\right.\nonumber\\[2mm]
&\times& \left. f[A^{g^{-1}},C]\right\}\nonumber\\[2mm]
&=& \frac{1}{V}
{\int}{\cal D}g{\cal D}A{\cal D}B{\cal D}{\bar{c}}{\cal D}c\,
\mbox{exp}
\left[iS[A]+i{\int}d^3x\sqrt{-G}G^{\alpha\beta}{\delta}_B
\left({\bar{c}}^a{\partial}_{\alpha}A_{\beta}^a\right)
\right]T_{\mu\nu}f[A^{g^{-1}},C]\, ,
\end{eqnarray}
where $T_{\mu\nu}$ is the canonical symmetric energy-momentum,
\begin{eqnarray}
T_{\mu\nu}&=&-\frac{2}{\sqrt{-G}}\frac{{\delta}S_{eff}}{{\delta}G^{\mu\nu}}
\nonumber\\[2mm]
&=&A_{\mu}^a{\partial}_{\nu}B^a+A_{\nu}^a{\partial}_{\mu}B^a
-{\partial}_{\mu}\bar{c}^a(D_{\nu}c)^a-{\partial}_{\nu}\bar{c}^a(
D_{\mu}c)^a\nonumber\\[2mm]
&-&G_{\mu\nu}\left[A_{\alpha}^a{\partial}^{\alpha}B^a-{\partial}_{\alpha}
\bar{c}^a(D^{\alpha}c)^a\right]\, ,\nonumber \\[2mm]
S_{eff}&=&S[A]+{\int}d^3x\sqrt{-G}G^{\alpha\beta}{\delta}_B\left(
\bar{c}^a{\partial}_{\alpha}A_{\beta}^a\right)\, .
\end{eqnarray}
It can be written as a BRST trivial form from a careful 
observation, 
\begin{eqnarray}
T_{\mu\nu}&=&{\delta}_B{\Theta}_{\mu\nu}\,,
\nonumber \\[2mm]
{\Theta}_{\mu\nu}&=&-{\partial}_{\mu}\bar{c}^aA_{\nu}^a-
{\partial}_{\nu}\bar{c}^aA_{\mu}^a-G_{\mu\nu}{\partial}^{\alpha}
\bar{c}^aA_{\alpha}^a\, .
\end{eqnarray}
So we can obtain that
\begin{eqnarray}
&-&\frac{2i}{\sqrt{-G}}\frac{{\delta}Z[C]}{\delta G^{\mu\nu}}
=-\frac{2i}{\sqrt{-G}}\frac{\delta F(C)}{\delta G^{\mu\nu}}{\bf 1}_{\rho}
\nonumber\\[2mm]
&=& {\int}{\cal D}A{\cal D}B{\cal D}{\bar{c}}{\cal D}c\,
\mbox{exp}\left\{iS[A]+i
{\int}d^3x\sqrt{-G}G^{\alpha\beta}{\delta}_B\left({\bar{c}}^a
{\partial}_{\alpha}A_{\beta}^a\right)
\right\}{\delta}_B{\Theta}_{\mu\nu}
\frac{1}{V}
{\int}{\cal D}gf[A^{g^{-1}},C]\nonumber\\[2mm]
&=&\langle 0|\,\left[\,\hat{Q}_B, \hat{\Theta}_{\mu\nu}\, \right]
\frac{1}{V}{\int}{\cal D}g
f[\hat{A}^{g^{-1}},C]\,|0 \rangle  
= \langle 0|\left[\, \hat{Q}_B, \hat{\Theta}_{\mu\nu}
\frac{1}{V}{\int}{\cal D}g
f[\hat{A}^{g^{-1}},C]\,\right]\,|0 \rangle  \nonumber\\[2mm]
&=&0\, ,
\label{eq:ge}
\end{eqnarray}
where $Q_B$ is the BRST charge corresponding to the BRST
transformation Eq.(\ref{eq8}) and the symbol hat ``\,$\hat{}$\,"
denotes an operator.  In the above, 
we have used the physical state condition in BRST
 quantization: $\hat{Q}_B|\mbox{phys} \rangle  =0$$\cite{ko}$ and the 
fact that $\displaystyle \frac{1}{V}
{\int}{\cal D}gf[A^{g^{-1}},C]$ is gauge invariant,
\begin{eqnarray}
\left[\,\hat{Q}_B, \frac{1}{V}
{\int}{\cal D}gf[\hat{A}^{g^{-1}},C]\,\right]=0
\end{eqnarray} 
as well as the explicit corollary observed by Witten$\cite{witten}$:
for two operators $\hat{A}$ and $\hat{B}$, if
$[\hat{Q}_B,\hat{A}]=0$, then $\hat{A}[\hat{Q}_B,\hat{B}]
=[\hat{Q}_B, \hat{A}\hat{B}]$.
Therefore from Eq.(\ref{eq:ge}), $Z[C]$ and $F[C]$ are generally covariant.

We shall now  justify the assumption that 
there is no metric independence for the path integral measure.  
This assumed property was very crucial in the proof of 
the general covariance of quantum holonomy.  
The justification can be made from two aspects.   First,  a metric 
dependence in the path integral measure means that under 
metric variation the path integral measure has a nontrivial Jacobian
factor.  According to Fujikawa$\cite{fu}$, this implies that the
theory would have a conformal anomaly, i.e.,  
the  trace of the energy-momentum 
$\langle{\Theta}_{\mu}^{\mu}\rangle$ does not vanish.  
 Since the  trace of the energy-momentum is proportional to the 
$\beta$-function of the theory$\cite{cd}$,  the existence of conformal 
anomaly would mean that the theory is not finite.  On the other hand, 
it has been proved that three-dimensional topological field theories
such as Chern-Simons$\cite{bc}$ and $BF$$\cite{lps}$ theories
are finite to any order, that is, the $\beta$-function and
anomalous dimension vanish identically.
This has also been verified by explicit computation in concrete 
regularization schemes to two-loop$\cite{gmr,csw,cc}$.
Therefore the conformal anomaly and hence the metric dependence
of the path integral measure should not exist. 

We could also approach the issue from the opposite direction and  
impose the conformal anomaly-free condition. Then the Jacobian 
associated with the metric variation would be trivial, which means that the 
correct path integration variables would not be the original fields 
${\Phi}=(A, B, c, \bar{c})$ but would be the appropriate tensorial densities
$\tilde{\Phi}$ given by 
 \begin{eqnarray}
\tilde{\Phi}_i=(-G)^{-w_i/2}{\Phi}_i \, ,
\end{eqnarray} 
where $G$ is the determinant of metric tensor $G_{\mu\nu}$, 
the subscript ``$i$" labels a specific field in $\Phi$ and 
$w_i$ is the weight associated with the field ${\Phi}_i$ whose 
value depends on the tensorial character of the corresponding
field.   Replacing  ${\Phi}_i$ by $\tilde{\Phi}_i$ as the path integral
variables and at the same time rewriting the action in terms of these
new variables, we could make the path integral measure metric 
independent. This procedure would  transfer the metric dependence 
of path integral measure to the effective action, and this would affect all 
the symmetries of the effective action such as BRST symmetry etc.  
 In Ref. \cite{cls}, it was shown that for a cohomological 
topological field theory, whose action can always be written as a 
BRST-trivial form, this line of reasoning can be used 
to define an invariant path integral measure so that   
the topological character of the theory is preserved.  
However, for topological field theories of the Chern-Simons type , 
whose action cannot be written as a BRST commutator, it is not 
clear how this technique can be applied.   
We intend to explore this problem in detail elsewhere.

Now we try to understand the result Eq.(\ref{eq6}) from the operator
viewpoint.  Since the global gauge symmetry 
is not affected by gauge fixing, the Noether current and 
charge corresponding to global gauge transformation are
 respectively, 
\begin{eqnarray}
j_{\mu}^a = -i\,\mbox{Tr}\frac{{\partial}{\cal L}}
{{\partial}{\partial}^{\mu}\Phi}[T^a,\Phi]\,, \qquad\qquad
Q^a = {\int}d^3 x j_0^a\, ,
\end{eqnarray}
where ${\Phi}{\equiv}(A, B, c, \bar{c})$.  After quantization, 
since  there is no anomaly in three dimensional gauge theory, 
we have
\begin{eqnarray}
{\partial}^{\mu}\langle \hat{j}_{\mu}^a \rangle  = 0\,,\qquad\qquad
\hat{Q}^a|0 \rangle  = 0\, .
\end{eqnarray}
The $\hat{Q}^a$ constitute the operator realization of 
the generators of the global gauge group, 
\begin{eqnarray}
[\hat{Q}^a,\hat{Q}^b]=if^{abc}\hat{Q}^c\,. 
\end{eqnarray}
Correspondingly,  
$U_{g^{'}}=\mbox{exp}[-i{\xi}^a\hat{Q}^a]$
 are global gauge group elements, ${\xi}^a$ are group parameters.  Under 
a global gauge transformation, the holonomy operator 
$f[\hat{A}, C]$ transforms as  
\begin{eqnarray}
f[\hat{A},C]{\longrightarrow}f[\hat{A},C]^{g'}
=U_{g^{'}}^{-1}f[\hat{A},C]U_{g^{'}}
={{\Omega}_{g^{'}}}^{-1}{f}[A,C]{\Omega}_{g^{'}}\, ,
\end{eqnarray}
where ${\Omega}_{g^{'}}= \mbox{exp}[-i{\xi}^aT^a]$ 
are the matrix representations of  group elements.  So we have
\begin{eqnarray}
Z[C]
&=&\langle 0|f[\hat{A},C]|0 \rangle 
=\langle 0|{U_{g^{'}}}^{-1}f[\hat{A},C]U_{g^{'}}|0 \rangle \nonumber\\[2mm]
&=&\langle 0|{{\Omega}_{g^{'}}}^{-1}f[{A},C]{\Omega}_{g^{'}}|0 \rangle  
={{\Omega}_{g^{'}}}^{-1}Z[C]{\Omega}_{g^{'}}\,.
\label{eq17}
\end{eqnarray}
$Z[C]$ commutes with every global gauge transformation, and 
from Schur's lemma we obtain Eq.(\ref{eq6}).

Finally let us consider the generalized quantum
holonomy defined on a multi-component link $L$, the disjoint union 
of $n$ simple knotted  contours $C_j$, $j=1,2\ldots n$. 
The holonomy operator is the tensor product of those defined on
each component $C_j$
\begin{eqnarray}
f[A, L]{\equiv}
P\mbox{exp}\left(i{\oint}_L A\right) =
{\bigotimes}_{j=1}^{n}\,P\mbox{exp}\,\left( i\,{\oint}_{C_j}A\right)\, . 
\end{eqnarray}

Using the same reasoning that was used to derive  Eq.(\ref{eq6}), 
we can see that  quantum holonomy defined a multi-component link 
commutes with any appropriately tensored generators of the gauge group,
\begin{eqnarray}
\left[\,{\bigotimes}_{i=1}^{n}T^a_{(i)}, Z[L]\, \right]=0\, ,
\qquad Z[L]=\langle f(A,L) \rangle  = 
\displaystyle \langle P\mbox{exp}\left(i{\oint}_L\,A \right)\rangle\, .  
\label{eqZL}
\end{eqnarray}
This shows that $Z[L]$ is a commutant of the universal 
enveloping algebra (of the Lie algebra of gauge group).   
The matrix representation of $Z[L]$ is not as simple as that of 
the quantum holonomy of a simple (one-component) knotted 
contour, however.   If  we now evaluate $Z[L]$ in the representation 
${\bigotimes}_{i=1}^{n}\rho_i$, the result will {\it not} be a polynomial 
times the $N$-dimensional matrix representation of the unity element, 
where $N$ is equal to the product of the dimensions of 
$\rho_i$, $i=1,\ldots,n$: 
\begin{eqnarray} 
N =  \prod_{i=1}^n N(\rho_i)\, .
\end{eqnarray}
  This  is because 
${\bigotimes}_{i=1}^{n}\rho_i$ is reducible.   Suppose this tensored 
representation decomposes as 
\begin{eqnarray}
{\bigotimes}_{i=1}^{n}\rho_i = 
{\bigoplus}_{j=1}^{m}\tau_j\, , \qquad
N= \sum_{j=1}^{m}N(\tau_j)\, ,
\label{decompose}
\end{eqnarray}
where each $\tau_j$  is irreducible, then it follows from Eq.(\ref{eqZL})  
that ${\bigotimes}_{i=1}^{n}\rho_i(Z[L])$  
will be a  diagonal matrix of dimension $N$, with its $N$ diagonal 
matrix elements being composed of possibly $m$ distinct polynomials, 
each polynomial repeating $N(\tau_j)$ times. 

The above is a general algebraic property of $Z[L]$, regardless of whether 
the theory is topological or not.   For a topological field theory, the 
polynomials in ${\bigotimes}_{i=1}^{n}\rho_i(Z[L])$   
carries additional topological information.  In the case of  
the Chern-Simons theory 
in three dimensions,  the polynomials will carry information pertaining 
to $L$ being a member of an isotopy class.   Let us take the traces of all 
the representations in ${\bigotimes}_{i=1}^{n}\rho_i$ 
except one, say that of $\rho_k$.   Then we obtain the counterpart of 
Eq.(\ref{eq6}), 
\begin{eqnarray} 
\bigotimes_{j\ne k}\mbox{Tr}_{\rho_j}
\left({\bigotimes}_{i=1}^{n}\rho_i(Z[L]) \right) 
 = F(L) {\bf 1}_{\rho_k}\, ,
\label{linkpoly}
\end{eqnarray}
where $F(L)$ is a polynomial of the $m$ polynomials in ${\bigotimes}_{i=1}^{n}\rho_i(Z[L])$; it is a polynomial for $L$.   
One might think that $F(L)$ would be labeled by $\rho_k$, but it has 
been shown$\cite{le2}$ that Eqs. (\ref{eqZL}) 
and (\ref{linkpoly}) are sufficient to prove that $F(L)$ is 
independent of the  choice of $\rho_k$.  Thus $F(L)$ is a uniquely 
defined eigenvalue of $Z[L]$ and is a link polynomial for $L$ on the 
set of representations $\{\rho_1, \rho_2,\cdots, \rho_n\}$.  

An explicit evaluation of $Z[L]$ by the right-hand-side of   
Eq.(\ref{eqZL}) is nontrivial.   This contrasts with the rather 
straightforward evaluation of the link polynomial $F(L)$ by 
algebraic means.   This method, based on long standing theorems by 
Alexander$\cite{alx}$ and Reidemeister$\cite{rei}$, 
begins by making a planar 
projection of a link, called a link diagram, 
in such a way that the projection is a network whose only nodes 
are either one of two kinds of  crossings -  overcrossing or 
undercrossing - each being a four-valent planar diagram.  
Topologically equivalent classes of links are classified according to 
isotopic classes of link diagrams. The connection to a gauge group 
either through a skein relation$\cite{jones}$, 
the braid group$\cite{wda}$ or directly$\cite{le2}$ 
is made by mapping the overcrossing (undercrossing, resp.) to 
(the inverse of, resp.) an 
invertible universal $R$-matrix, which is a construct 
(more specifically, a second rank tensor product) that 
exists in universal enveloping algebras of Lie algebras such as 
the Lie algebra of $SU(N)$.  

 In Ref. \cite{wet},  the emergence of an $R$-matrix in  
Wilson loops of the three-dimensional Chern-Simons theory was 
investigated.   It was shown that by choosing 
a special gauge - the almost axial gauge - and working in the  
space-time manifold $S^1{\times}R^2$,
one can use the technique of standard perturbation theory to reveal
the assignment of an $R$-matrix to the crossing on a link diagram.  
Since the trace of the Wilson loop was actually not taken in Ref. \cite{wet}, 
its conclusions more appropriately apply  
to quantum holonomy.  However, because the object of 
investigation, the link invariant, is a nonperturbative property 
of the Chern-Simons theory, the conclusion is somewhat 
clouded through a lack accuracy owing to the nature of the 
perturbation method.    Perhaps a direct 
nonperturbative evaluation of Eq.(\ref{eqZL}),  
such by lattice gauge theory, is called for.   
\vspace{3ex}

\acknowledgments 
HCL is partially supported by grant 
85-2112-M-008-011 from the National Science Council, ROC 
and WFC is grateful to ICSC-World Laboratory,  Switzerland for financial 
support.

\end{document}